\documentclass[twocolumn,apl,amsmath,amssymb,showpacs,superscriptaddress]{revtex4-2}
\usepackage{epsf}      
\usepackage{graphicx}
\usepackage{color}
\usepackage{soul}
\usepackage[T1]{fontenc}
\usepackage{gensymb}
\usepackage{sidecap}
\usepackage{amsmath}
\DeclareUnicodeCharacter{202F}{}
\usepackage{mathtools}
\usepackage{float}
\usepackage[hidelinks,colorlinks=true,linkcolor=blue,citecolor=blue]{hyperref}

\begin{document}
\title{Physical and Dielectric Properties of Polycrystalline LaV$_{0.5}$Nb$_{0.5}$O$_4$}

\author{Ashok Kumar}
\affiliation{Department of Applied Physics, Delhi Technological University, Delhi-110042, India}
\affiliation{Department of Physics, Atma Ram Sanatan Dharma College, University of Delhi, New Delhi-110021, India}
\author{Simranjot K. Sapra}
\affiliation{Department of Physics, Indian Institute of Technology Delhi, Hauz Khas, New Delhi-110016, India}
\author{Ramcharan Meena}
\affiliation{Material Science Division, Inter-University Accelerator Center, Aruna Asaf Ali Road, New Delhi-110067, India}
\author{Vinod Singh}
\affiliation{Department of Applied Physics, Delhi Technological University, Delhi-110042, India}
\author{Anita Dhaka}
\affiliation{Department of Physics, Hindu College, University of Delhi, New Delhi-110007, India}
\author{Rajendra S. Dhaka}
\email{Corresponding author: rsdhaka@physics.iitd.ac.in}
\affiliation{Department of Physics, Indian Institute of Technology Delhi, Hauz Khas, New Delhi-110016, India}

\date{\today}      

\begin{abstract}
We report a detailed investigation of the structural, electronic, vibrational, and dielectric properties of polycrystalline LaV$_{0.5}$Nb$_{0.5}$O$_4$ samples, prepared at two  sintering temperatures (1000\degree C and 1250\degree C). The introduction of Nb$^{5+}$ at the V$^{5+}$ site leads to notable structural and vibrational changes, which can be attributed to their isoelectronic nature and the comparatively larger ionic radius of Nb$^{5+}$. The Rietveld refinement of the X-ray diffraction patterns confirms a coexistence of monoclinic ($P$2$_{1}$/$n$) and scheelite-type tetragonal ($I$4$_{1}$/$a$) phases; for example, with a fraction of 4\% and 96\% for the sample annealed at 1250\degree C. The particle morphology has altered from spherical (1000\degree C) to irregular-shaped (1250\degree C) as a result of increase in annealing temperature. The Raman spectroscopy, Fourier Transform Infrared spectroscopy and X-ray Photoemission Spectroscopy have been used to understand the vibrational and electronic properties. An optical band gap of 2.7~eV for the sample sintered at 1250\degree C is calculated using Ultraviolet-vis diffuse reflectance spectroscopy measurements. The dielectric studies shows the higher dielectric permittivity ($\epsilon$$_{r}$) and lower dielectric loss for the sample annealed at 1250\degree C.
 
\end{abstract}

\maketitle

\section{\noindent ~Introduction}

In last few decades, rare-earth orthovanadates have emerged as a significant class of functional materials with broad environmental and industry applications spanning as luminescent materials, catalysis, solid oxide fuel cells, battery electrodes, thermometers and photocatalysts \cite{YiJAL17, XiaJAP00, Michalska2021, Xu_CEComm_2011, HimanshuPRB21}. The orthovanadates with general formula RVO$_4$, where R is a rare-earth ion, such as La$^{3+}$, Gd$^{3+}$, Y$^{3+}$ etc., represent a versatile family of materials having the VO$_4^{3-}$ tetrahedral units, which serve as the fundamental building blocks \cite{Shwetha2015, Errandonea2020, Thakur2024, Ashok_Jalcom_23}. It allows the rare earth cations to form a crystal field, which upon perturbation, results in energy transition and phenomena of photoluminescence \cite{Singh_Optik_2021}. In the family, the lanthanum based orthovanadates, i.e., LaVO$_{4}$ has been considered most interesting due to its exceptional optical, electronic and structural properties \cite{Michalska2021} as well as used for targeted alpha therapy in cancer treatment through radionuclide encapsulation using $^{223}$Ra to form La($^{223}$Ra)VO$_{4}$ \cite{Gonzalez_JNT_21}. Interestingly, the LaVO$_4$ exhibits peculiar structural polymorphism, crystallizing primarily in two structure types: the tetragonal (t-) zircon-type (space group I4$_1$/amd [141]) and the monoclinic (m-) monazite-type (space group P2$_1$/n [14]) \cite{Jia_JPCB_2005, Cheng_OM_2015}. In the monoclinic monazite structure, La$^{3+}$ ions are coordinated by nine oxygen atoms forming edge sharing LaO$_9$ polyhedra and connected with distorted VO$_4$ tetrahedra with four different V--O bond lengths. However, in the tetragonal zircon structure, La$^{3+}$ ions are eight-fold coordinated with the equal V--O bond lengths and linked together through chains along the $c$--axis. Importantly, the LaVO$_4$ shows stability in monoclinic monazite structure at room temperature with La$^{3+}$ ion's high tendency to crystallize in nine-coordination as compared to the eight coordination in tetragonal zircon structure \cite{Cheng_OM_2015}. This is mainly because the phase stability is dependent on the cation radius and La$^{3+}$ has the largest ionic radius (1.16 \AA) as compared to other rare earth cations. The tetragonal (zircon-type) phase of LaVO$_4$ is metastable at room temperature, but can be synthesized through solution-based methods such as hydrothermal  \cite{Cheng_OM_2015, Fan_JPCB_2006}. Interestingly, the tetragonal structure exhibits superior luminescence properties due to the presence of four identical V--O bonds (angle $\approx 153^\circ$), which facilitate more efficient energy transfer \cite{Park_PBCM_2010}. In contrast, the monoclinic monozite phase has smaller bond angles, resulting in less effective energy transfer \cite{Park_PBCM_2010}. The LaVO$_4$ is a self-activated ABO$_4$ type host compound, emitting broad blue emission from 380 nm to 450 nm region with maxima at 430 nm upon UV light excitation. This corresponds to a transition from two close lying $^1$A$_1$($^1$E) and $^1$E$_1$($^1$T$_2$) excited states to $^1$A$_2$($^1$T$_1$), ground state of (VO$_4$)$^{3-}$ ion \cite{Rai_SSS_2022}. Here, the Lanthanum is significantly more abundant and less expensive than Yttrium, making LaVO$_4$-based materials economically attractive alternatives to the widely used YVO$_4$ phosphors \cite{Wang_JALCOM_2023}. This cost advantage is particularly important for large-scale applications in lighting and display technologies.

Notably, the structural diversity enables tunable optical and electronic properties through composition and external conditions such as pH, temperature and pressure \cite{Gonzalez_JNT_21, Cheng_OM_2015, Yang2018}. Therefore, substitution at V$^{5+}$ site in LaVO$_4$ has been explored through various dopants, such as Mn$^{4+}$ where LaV$_{1-x}$Mn$_x$O$_{4-\delta}$ (0 $\leq$ x $\leq$ 1) mixed oxides were synthesized in single-phase with reproducible redox behavior (LaVO$_{4}$ to LaVO$_3$), which indicate that the Mn doping lowered reduction/oxidation temperatures and enhanced oxygen vacancy formation and mobility for catalytic performance \cite{Varma_ACA_2001}. Interestingly, substituting low-cost Nb$^{5+}$ at the V$^{5+}$ site in LaVO$_4$ motivated by the intriguing structural and optical behavior arising from the isoelectronic character (d$^0$ electronic configuration) and larger ionic radii of Nb$^{5+}$ as compared to V$^{5+}$ \cite{HimanshuPRB21, Ashok_Jalcom_23, Thakur2024}. It has also been well-demonstrated that even small amounts of Nb$^{5+}$ substitution ($x \geq 0.05$) can induce the formation of the scheelite-tetragonal phase (I4$_1$/a) alongside the monoclinic phase at room temperature \cite{HimanshuPRB21}. The presence of the tetragonal phase significantly enhances photoluminescence properties and the structural transformation can be controlled by varying the Nb concentration \cite{HimanshuPRB21, Ashok_Jalcom_23, Thakur2024}. Meanwhile, the end composition in the series, i.e., the LaNbO$_4$ is a rare-earth niobate and structurally transforms from monoclinic fergusonite ($I$2/$a$) to tetragonal scheelite ($I$4$_1$/a) at 495$\degree$C \cite{TakeiJCG77}. On the other hand, Ding {\it et al.} reported the LaNbO$_4$ in monoclinic phase, which exhibits interesting luminiscent features with fluorescence lifetime of 5.51~$\mu$s \cite{Ding2018}. In a similar fashion, when V$^{5+}$ ion is substituted at the Nb$^{5+}$, the composition induced structural transition is observed \cite{AldredML83}. Upon excitation with a UV or X-ray source, charge transfer from O 2$p$ to Nb 4$d$ band in LaNbO$_4$ yielded a blue luminescent material, as demonstrated by Blasse {\it et al.} in ref.~\cite{BlasseCPL90}. However, the composition LaV$_{0.5}$Nb$_{0.5}$O$_4$ remains largely unexplored and presents significant scope for detailed investigation particularly to see the effect of sintering temperature during the synthesis process as well as temperature dependent dielectric properties \cite{HimanshuPRB21, Ashok_Jalcom_23, Thakur2024}.  

Therefore, the primary aim of this work is to study polycrystalline lanthanum orthovanadate with niobium substitution, specifically targeting the composition LaV$_{1-x}$Nb$_x$O$_4$ where $x$ equals 0.5, utilizing the solid-state method and investigate sintering temperature induced phase transformation and understand the dielectric properties. Interestingly, the crystal structure and morphology of the orthovanadates is strongly dependent upon the synthesis approach, annealing temperature and concentration of active ions. Herein, the LaV$_{0.5}$Nb$_{0.5}$O$_4$ sample crystallize in two distinct, monoclinic monazite type and tetragonal zirconite type phases in different fractions, as confirmed by the Rietveld refinement of X-ray diffraction (XRD) patterns. The scanning electron microscopy (SEM) measurements unveil the morphological variations with the sintering temperature. The Raman spectroscopy and Fourier transform infrared spectroscopy (FTIR) measurements identify the presence of vibrational bonds in the prepared samples. The energy band gap values have been estimated through ultraviolet-visible diffuse reflectance spectroscopy measurements. The core-level X-ray photoemission spectroscopy studies depict the electronic structure and oxidation state of the constituent elements in both the samples. The microstructure has been successfully confirmed by the high-resolution transmission electron microscopy measurements. The dielectric measurements  reveal the dielectric permittivity ($\epsilon$$_{r}$) and dielectric loss in the temperature range of 100 ~K to 400~K for the samples annealed at 1000$^{\circ}$C and 1250$^{\circ}$C. The {\it a.c.} conductivity studies are performed to understand the charge carrier dynamics with variable temperature and frequencies. 

\section{\noindent ~Experimental}

{\bf \noindent Materials Preparation}: The LaV$_{0.5}$Nb$_{0.5}$O$_4$ bulk sample was synthesized via a conventional solid-state reaction route. Prior to synthesis, La$_2$O$_3$ was pre-heated at 900\degree C for 6~h with a ramp rate of 4~$^\circ$C~min$^{-1}$ in order to remove the adsorbed moisture. The calcined La$_2$O$_3$ was quenched to room temperature and weighed immediately to prevent re-hydration. High-purity La$_2$O$_3$, V$_2$O$_5$, and Nb$_2$O$_5$ were mixed in the stoichiometric ratio to obtain a 2~g final yield of the powder. The precursors were thoroughly mixed and ground in an agate mortar and pestle for 6~hr to ensure homogeneous mixing. The resulting powder was pelletized under a uniaxial pressure of 1500~kg~cm$^{-2}$, followed by calcination at 1000\degree C for 17~hr with a heating rate of 5\degree C~min$^{-1}$. The calcined product was reground for approximately 5~hr to ensure homogeneity and subsequently re-pelletized under the same applied pressure (1500~kg~cm$^{-2}$). The final sintering was carried out at 1250$^\circ$C for 13~hr with a ramp rate of 5\degree C~min$^{-1}$. The LaV$_{0.5}$Nb$_{0.5}$O$_4$ samples annealed at 1000\degree C and 1250\degree C are abbreviated as LVNO-1000 and LVNO-1250, respectively.

{\bf \noindent Physical and Dielectric characterizations}: The X-ray diffraction (XRD) measurements are carried out using a Panalytical X'Pert$^{3}$ diffractometer equipped with Cu K$_{\alpha}$ radiation ($\lambda=$ 1.5406~\AA). The Rietveld refinement of the powder XRD patterns is performed using the FullProf software, with the background fitted using linear interpolation between the data points. The morphology of the bulk samples is examined using a field-emission scanning electron microscope (FE-SEM, Technai Magna LMU). The elemental mappings are probed through the EDX detector system (EDAX AMETEK, Octane Elite Super EDS). The Fourier-transform infrared spectroscopy (FTIR, Thermo Nicolet IS-50) is performed in Attenuated Total Reflectance (ATR) mode over the spectral range of 400--4000~cm$^{-1}$ to analyse vibrational modes. The X-ray photoelectron spectroscopy (XPS, Kratos AXIS Supra) is employed to investigate the valence state using monochromatic Al K$_{\alpha}$ radiation (1486.6~eV) as the excitation source, with a pass energy of 20~eV and a step size of 0.1~eV. The spectral fitting is performed using FitYk software with Voigt function (Shape factor = 0.6), and the C~1$s$ peak at 284.6~eV was used as the reference for calibration. The microstructure of the bulk samples is examined using transmission electron microscopy (TEM, JEM-ARM200F NEOARM). The Raman spectroscopy is performed using a Renishaw inVia confocal Raman microscope equipped with a 532~nm laser, a 2400 lines/mm grating, and a laser power of 10 mW to investigate the vibrational modes of the samples. The UV-vis (Ultraviolet-visible) diffuse reflectance spectroscopy (DRS) measurements are conducted using a Perkin Elmer's Lambda 1050 model in the spectral range of 300--2500~nm to measure the absorption edge energy of the samples using the Barium Sulfate standard. The pellets were polished and painted with sliver paste on both the sides for the dielectric measurements. The dielectric studies are done in the temperature range of 90--400 K by varying the frequency from 100 Hz--2 MHz using the Agilent LCR meter (model E4980A). The temperature is varied using the Lakeshore temperature controller (model-340) and temperature was measured using the Pt-100 sensor. The measurements are performed in the vacuum of the order 10$^{-2}$ mbar by placing the sample holder in liquid nitrogen Dewar.

\section{\noindent ~Results and discussion}

\begin{figure}
	\includegraphics[width=\linewidth]{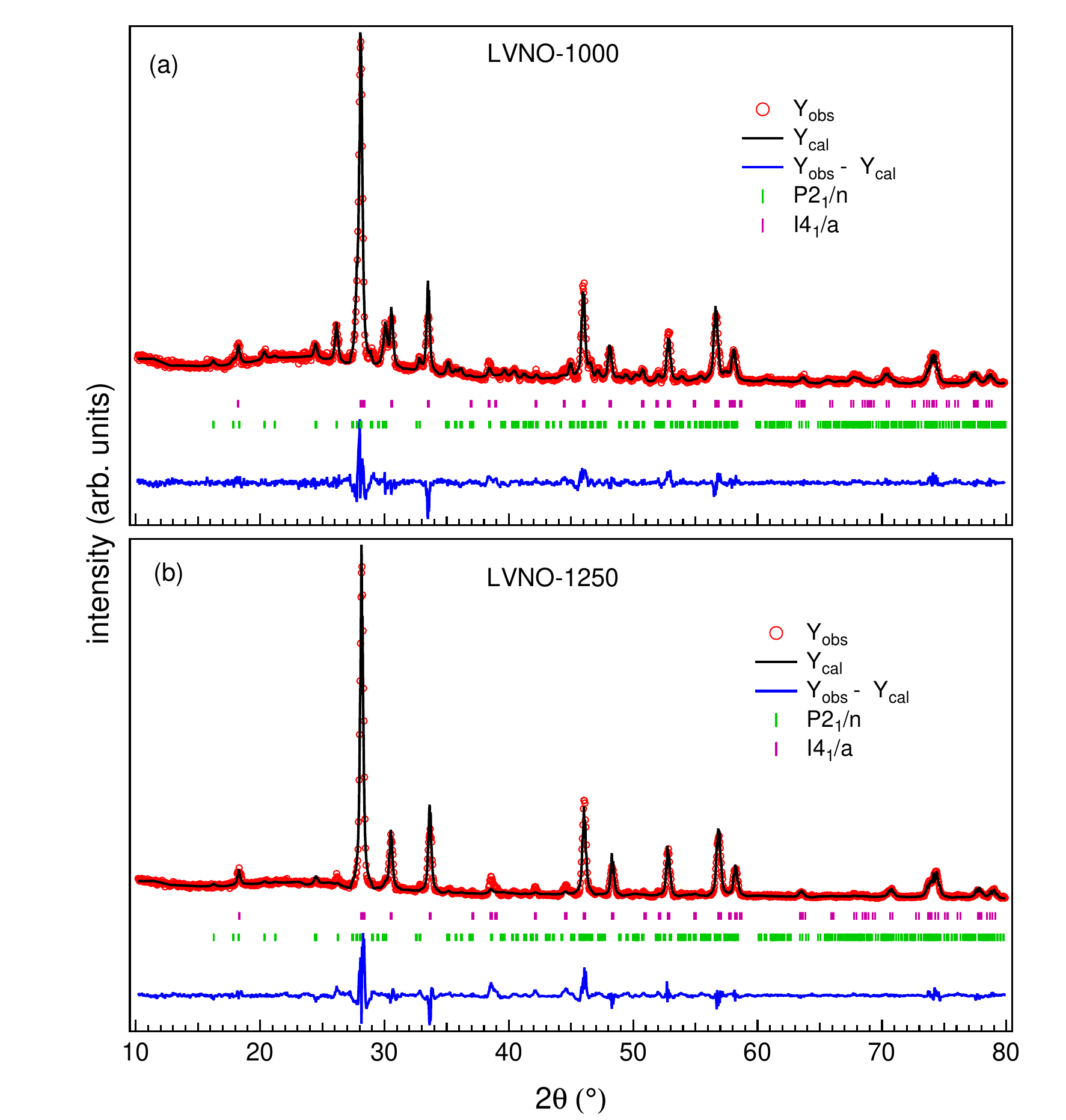} 
	\caption{The room temperature XRD patterns with Rietveld refinement of (a) LVNO-1000 and (b) LVNO-1250 samples. The open red circles, black solid line, and blue solid line exhibits the experimental, calculated, and the difference between experimental and calculated pattern, respectively. The vertical green and purple markers show the Bragg positions corresponding to the $P$2$_{1}$/$n$ and $I$4$_{1}$/$a$ space groups. } 
	\label{Fig1}
\end{figure}

The room temperature Rietveld refined XRD patterns of as-prepared polycrystalline LaV$_{0.5}$Nb$_{0.5}$O$_4$ samples, annealed at 1000\degree C and 1250\degree C are depicted in Figs.~\ref{Fig1}(a, b), respectively. It is observed that the measured XRD pattern consists of a mixture of monoclinic ($P$2$_{1}$/$n$) and scheelite-tetragonal ($I$4$_{1}$/$a$) phases in varying proportions, rather than a single $P$2$_{1}$/$n$ phase \cite{AldredML83}. The structural phase transformation in the LVNO sample is caused by the Nb$^{5+}$ substitution at the V$^{5+}$ site in a tetrahedral coordination, because the ionic radius of Nb$^{5+}$ (0.48~\AA)  is larger than that of V$^{5+}$ (0.355~\AA) \cite{Shannon}. It is also reported in the literature that end members, the LaVO$_4$ and LaNbO$_4$ co-exist in the monoclinic monazite ($P$2$_{1}$/$n$) and fergusonite monoclinic ($I$2/$a$) phases, respectively, at room temperature \cite{Guo_IC_2017, Huse_JSSC_2012}. Consequently, it is anticipated that Nb$^{5+}$ substitution in LaVO$_4$ will induce a phase transformation from $P$2$_{1}$/$n$ to $I$2/$a$. In the LaVO$_{4}$ sample, synthesized at 1000\degree C, the XRD pattern is well fitted using monoclinic ($P$2$_{1}$/$n$) and scheelite-tetragonal ($I$4$_{1}$/$a$). The phase exhibits diffraction peaks characteristic of a crystalline solid; however, the peaks appear relatively broad and of lower intensity compared to those of the sample sintered at 1250\degree C. This broadening indicates a reduced degree of crystallinity and/or smaller crystallite size in the LVNO-1000 sample. In contrast, the LVNO-1250 sample displays sharp and intense diffraction peaks, signifying enhanced crystallinity and larger crystallite dimensions, as shown in Fig.~\ref{Fig1}(b). The higher synthesis temperature likely promotes more complete crystallization and grain growth. The diffraction peaks for both the samples sintered at 1000\degree C and 1250\degree C correspond to two structural phases: the scheelite-type tetragonal phase ($I$4$_{1}$/$a$) and the monoclinic phase ($P$2$_{1}$/$n$).  

\begin{table}[ht]
	\centering
	\caption{The structural and atomic parameters obtained from Rietveld refinement of the sample sintered at 1000\degree C.}
	
	\footnotesize
	\setlength{\tabcolsep}{8pt}
	\renewcommand{\arraystretch}{1.2}

	\begin{tabular}{p{0.8cm} p{1cm} p{1cm}p{1cm} p{1cm} p{0.5cm}}
		\hline
		\multicolumn{6}{c}{\text{Tetragonal Phase:} $a=b=5.348$ \AA, $c=11.694$ \AA} \\
		\hline
		Atom & x & y & z & Occ. & Site \\
		\hline
		Nb & 0.0000 & 0.0000 & 0.0000 & 1.00 & 36f \\
		O  & 0.2635 & 0.8049 & 0.1435 & 1.00 & 36f \\
		La & 0.0000 & 0.0000 & 0.5000 & 1.00 & 6b \\
		\hline
		\hline

		\multicolumn{6}{c}{\text{Monoclinic Phase:} $a=7.054$ \AA, $b=7.278$ \AA, $c=6.731$ \AA
		} \\
		\multicolumn{6}{c}{ $\beta=104.909^\circ$
		}\\
		\hline
		Atom & x & y & z & Occ. & Site \\
		\hline
		V  & 0.3071 & 0.1657 & 0.6066 & 1.00 & 36f \\
		O1 & 0.2438 & 0.0155 & 0.4594 & 1.00 & 36f \\
		O2 & 0.3676 & 0.3134 & 0.5676 & 1.00 & 36f \\
		O3 & 0.4553 & 0.0593 & 0.8394 & 1.00 & 6b \\
		O4 & 0.1324 & 0.2049 & 0.7648 & 1.00 & 18e \\
		La & 0.2723 & 0.1613 & 0.1128 & 1.00 & 18e \\
		\hline
	\end{tabular}
	\label{xrd-1000}
\end{table}

\begin{figure*}
	\includegraphics[width=\linewidth]{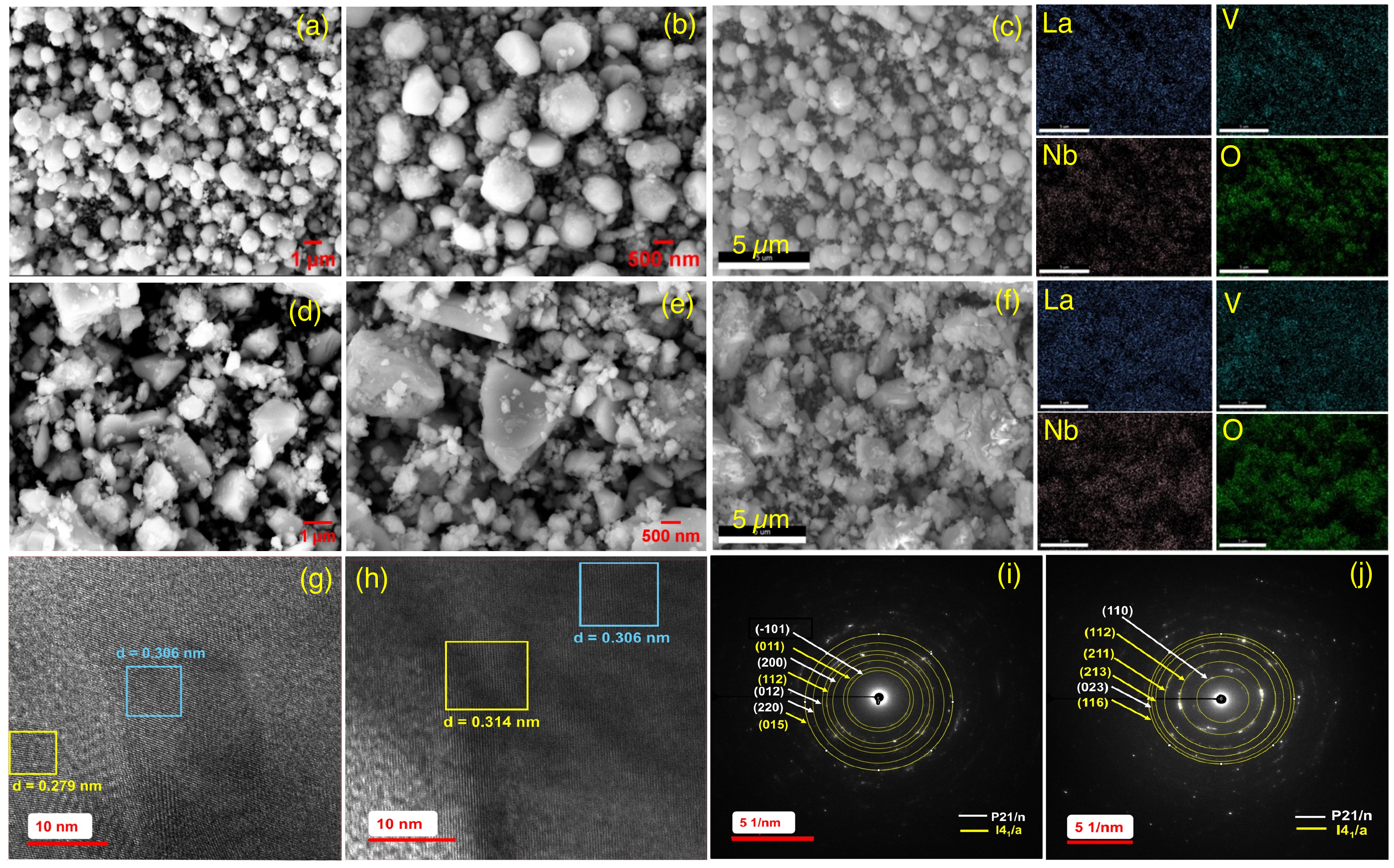} 
	\caption{The FE-SEM images of the LVNO-1000 sample at (a) 1$\mu$m, (b) 500~nm and (c) probed region of scan (at 5$\mu$m) with the corresponding elemental mappings of all elements; the FE-SEM images of the LVNO-1250 sample at (d) 1$\mu$m, (e) 500~nm and (f) probed region of scan (at 5$\mu$m) with the corresponding elemental mappings of all elements; the HR-TEM images of (g) sintered at 1000\degree C and (h) sintered at 1250\degree C with the marked $d$-spacings across the selected region; the SAED patterns for the LVNO-1000 (i) and LVNO-1250 (j), respectively. } 
	\label{Fig2}
\end{figure*}

\begin{table}[ht]
	\centering
	\caption{The Structural and atomic parameters obtained from Rietveld refinement of LVNO sintered at 1250\degree C.}
	
	\setlength{\tabcolsep}{8pt}
	\renewcommand{\arraystretch}{1.2}
	
	\begin{tabular}{p{0.8cm} p{1cm} p{1cm} p{1cm} p{1cm} p{0.5cm}}
		\hline
		\multicolumn{6}{c}{\text{Tetragonal Phase:} $a=b=5.327$ \AA, $c=11.716$ \AA} \\
		\hline
		Atom & x & y & z & Occ. & Site \\
		\hline
		Nb  & 0.0000 & 0.0000 & 0.0000 & 1.00 & 36f \\
		O   & 0.2500 & 0.7682 & 0.1576 & 1.00 & 36f \\
		La  & 0.0000 & 0.0000 & 0.5000 & 1.00 & 6b  \\
		\hline
		\hline
		
		\multicolumn{6}{c}{\text{Monoclinic Phase:} $a=7.027$ \AA, $b=7.293$ \AA, $c=6.732$ \AA} \\
		\multicolumn{6}{c}{ $\beta=104.947^\circ$
		}\\
		\hline
		Atom & x & y & z & Occ. & Site \\
		\hline
		V   & 0.3024 & 0.1699 & 0.5733 & 1.00 & 36f \\
		O1  & 0.1951 & 0.0846 & 0.4725 & 1.00 & 36f \\
		O2  & 0.5818 & 0.3642 & 0.5407 & 1.04 & 36f \\
		O3  & 0.8340 & 0.1151 & 0.7265 & 0.95 & 6b  \\
		O4  & 0.0960 & 0.6676 & 0.2146 & 0.76 & 18e \\
		La  & 1.6958 & 0.2865 & 0.1337 & 1.00 & 18e \\
		\hline
	\end{tabular}
	\label{xrd-1250}
\end{table}

The lattice parameters and wyckoff positions of LVNO-1000 and LVNO-1250 samples are summarized in Table~\ref{xrd-1000} and Table~\ref{xrd-1250}, respectively. It is observed that angle $\beta$ has slightly increased with increase in sintering temperature, as a result of increase in tetragonal phase fraction. The quantification of the relative phase fractions for both the samples has also been investigated through Rietveld refinement. The phase fractions of 51\% tetragonal ($I$4$_{1}$/$a$) and 49\% monoclinic ($P$2$_{1}$/$n$) are calculated for the LVNO-1000 sample. On the other hand, the LVNO-1250 sample exhibits a dominant tetragonal fraction of 96\% and a monoclinic fraction of 4\%. The predominance of the tetragonal phase at the higher temperature is consistent with its thermodynamic stability and structure becomes more homogeneous through completion of solid-state reaction. This is consistent with the previous work, where sintering of LaV$_{1-x}$Nb$_x$O$_4$ ($x$ = 0.1, 0.2) at 1450$\degree$C promoted the stabilization of scheelite-tetragonal phase \cite{HimanshuPRB21}. In addition, Xie {\it et al.} reported the mechanism of formation of tetragonal phase as a result of intermediate complex V$_x$O$_y$$^{(2y-5x)-}$, which forces La$^{3+}$ to coordinate in fewer coordination value (here, 8) due to strong stearic hindrance \cite{Xie_JALCOM_2012}. This increment in tetragonal phase fraction is better for superior luminescent characteristics \cite{Okram_IC_2014}.

The FE-SEM images of the LVNO samples, sintered at 1000\degree C and 1250\degree C are presented in Figs.~\ref{Fig2}(a--f). The LVNO-1000 sample shows an average particle size of approximately 0.8~\textmu m. In contrast, the LVNO-1250 sample exhibits a non-uniform particle size distribution ranging from 0.2~\textmu m to several micrometers, as depicted in Fig.~\ref{Fig2}(b). The grain size has increased with the increase in sintering temperature, as the shape changing from spherical (at 1000\degree C) to irregular-shaped (at 1250\degree C) particles. The morphological difference along with the visible textural changes, such as inevitable agglomeration in the LVNO-1250 sample, provide additional evidence of the temperature-induced phase transformation. The grain boundaries diffuse, smaller particles rearrange to large sized particles and coalescence occurs as a result of high temperature annealing \cite{Li_CRev_2022}. Furthermore, the uniform distribution of elements in LVNO-1000 and LVNO-1250 samples can be clearly observed in Figs.~\ref{Fig2}(c, f). The HR-TEM analysis of the LVNO-1000 sample clearly reveals its biphasic nature, with distinct coexisting lattice fringes exhibiting $d$-spacings of approximately 0.28~nm and 0.3~nm, depicted in Fig.~\ref{Fig2}(g). The indexed (h, k, l) planes that relate to $P$2$_{1}$/$n$ and $I$4$_{1}$/$a$ are colored in white and yellow, respectively, see Figs.~\ref{Fig2}(g, h). The corresponding SAED pattern further confirms the mixed phase structural arrangement, where the reflections (-101), (200), (012) and (220) belong to the monoclinic phase and (011), (112) and (015) belong to the tetragonal phase, see Figs.~\ref{Fig2}(i, j). For the LVNO-1250 sample, the HR-TEM images indicate a dominant tetragonal phase [see Fig.~\ref{Fig2}(h)], with measured $d$-spacings shifting to around 0.31~nm, which are characteristic of the major tetragonal reflections. The SAED pattern for this sample is overwhelmingly dominated by diffraction spots consistent with the I4$_1$/a phase, including the (112), (211), and (213) reflections [see Fig.~\ref{Fig2}(j)], providing strong evidence that higher sintering temperature promotes the stabilization and predominance of the tetragonal phase. This is in agreement with the refined X-ray diffraction analysis, presented in Figs.~\ref{Fig2}(a, b).

\begin{figure}[h]
	\includegraphics[width=\linewidth]{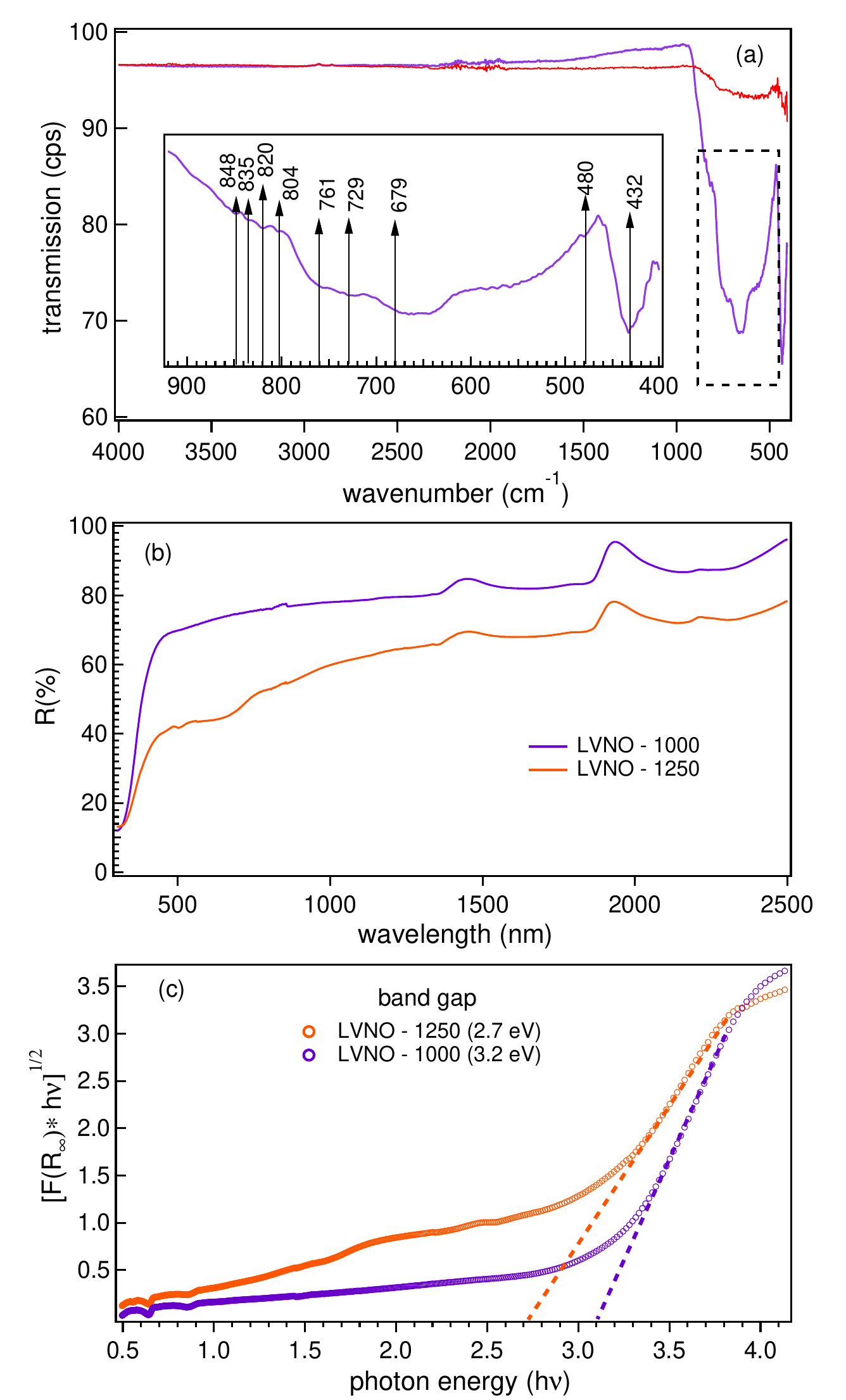} 
	\caption {(a) The FTIR spectra, (b) the UV-vis DRS plots and (c) the Kubelka-Munk plots obtained from the DRS data for the LVNO-1000 and LVNO-1250 samples, respectively.} 
	\label{Fig3}
\end{figure}

In order to confirm the structure and possibility of infrared active modes of the LaV$_{0.5}$Nb$_{0.5}$O$_4$ sample, the ATR-FTIR measurements have been performed, as depicted in Fig.~\ref{Fig3}(a), with the closer view shown in the inset. The observed well-resolved bands are marked with the black vertical arrows [inset of Fig.~\ref{Fig3}(a)] and are in close agreement with ref.~\cite{HimanshuPRB21}. Herein, a strong and distinct band is observed at 432 cm$^{-1}$, with a weak band at 478 cm$^{-1}$, which are linked to the $\nu$$_{4}$ bending vibrations of the (V/Nb)O$_{4}$$^{3-}$ tetrahedra \cite{Michalska2021}. Similarly, wide bands are observed between 600--900 cm$^{-1}$, where a peak at 804 cm$^{-1}$ corresponds to the antisymmetric stretching vibrations of VO$_{4}$$^{3-}$ tetrahedra. Meanwhile, the peaks at 820, 835 and 848 correspond to the $\nu$$_{3}$ stretching vibrations of VO$_{4}$$^{3-}$ tetrahedra \cite{Liu_MCP_2009} and show a small shift in the peaks in comparison to pristine LaVO$_{4}$ sample \cite{HimanshuPRB21}. In addition, the peaks at 679, 729 and 761 cm$^{-1}$ have been originated with the incorporation of Nb at V site and are absent in the LaVO$_{4}$ sample. These features can be correlated with the additional symmetry modes, arising from the mixed scheelite-tetragonal phase, as proven in X-ray diffraction measurements. However, the ATR-FTIR spectra for LVNO-1250 sample shows less intense band around 417 cm$^{-1}$, while the peaks in the range of 600-900 cm$^{-1}$ appear as a single broad band, depicted in Fig.~\ref{Fig3}(a). It can be inferred that IR activity gets reduced with increasing temperatures due to increase in thermal energy and as a result, intensity of specific vibrational bands are reduced.

\begin{figure*}
	\includegraphics[width=\linewidth]{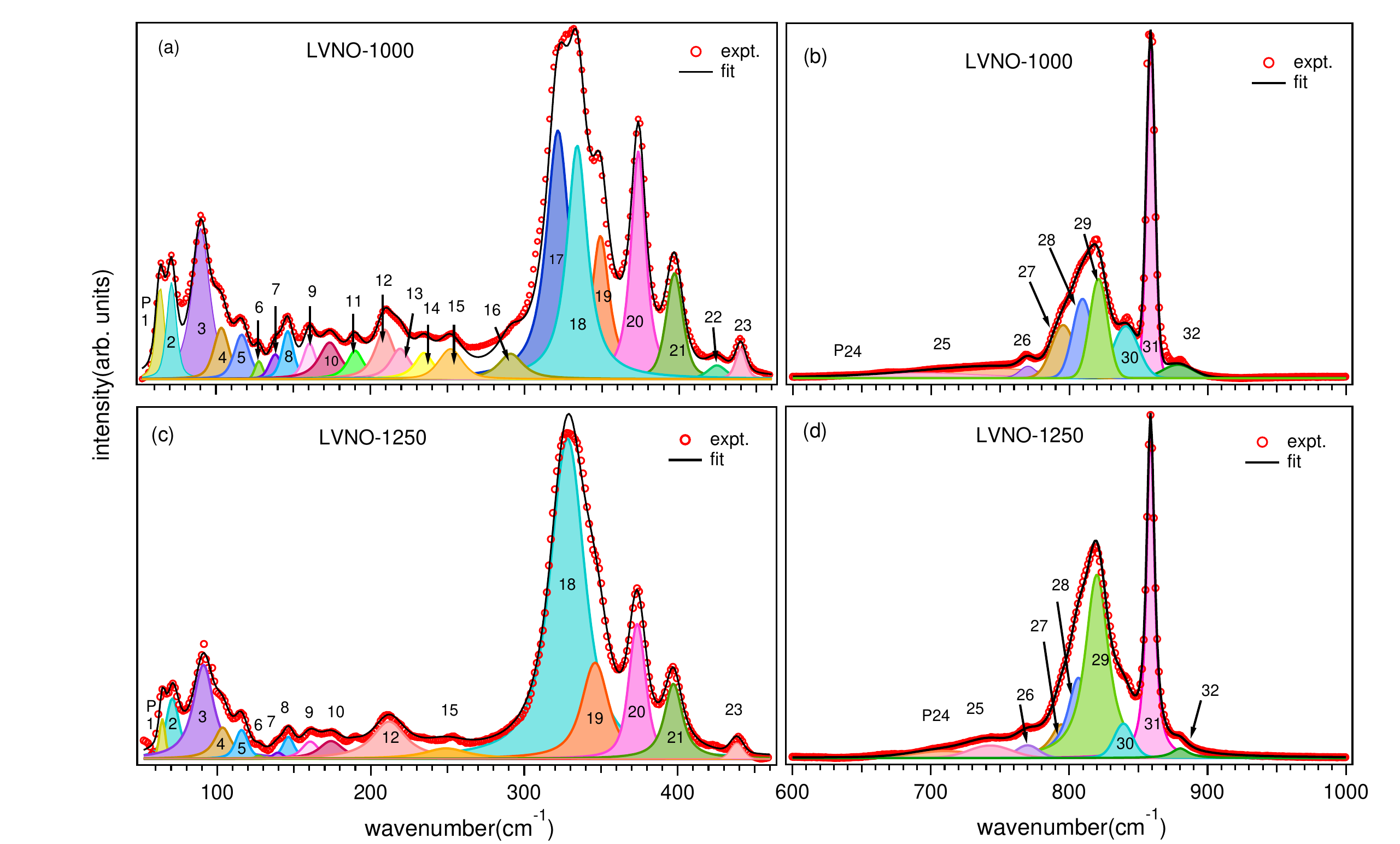} 
	\caption {The room temperature Raman spectra of LVNO sample, (a, b) sintered at 1000\degree C and (c, d) for sample sintered at 1250$\degree$C in the low and middle wavenumber regions, respectively; fitted using the Voigt peak function and solid thick black line presents the total fit of the measured spectra.} 
	\label{Fig4}
\end{figure*}

The UV-vis DRS measurements are conducted to understand the absorption edge energy of the synthesised LVNO samples, sintered at 1000\degree C and 1250\degree C. The reflectance spectra [see Fig.~\ref{Fig3}(b)] observed for both the samples is converted to Tauc plot using the Tauc method \cite{Tauc_PSS_1966}. The diffuse reflectance spectra (DRS) are employed to calculate the Kubelka-Munk (K-M) function [F(R$_{\infty}$)], shown in Fig.~\ref{Fig3}(c), which can be treated in terms of diffuse semi-infinite reflectance (R$_{\infty}$ = R$_{sample}$/R$_{standard}$) as written below \cite{Kubelka_Phy_1931}:
\begin{equation}
	F(R_\infty) = \frac{(1-R_\infty)^2}{2R_\infty}
\end{equation}
The band gap of the LVNO samples can be obtained from the absorption threshold energy, which is the $x-$axis intercept of the Tauc plot [see Fig.~\ref{Fig3}(c)] and related as~\cite{Lopez_JSST_2012}:
\begin{equation}
	(\alpha h\nu) = k (h\nu - E_g)^\eta
\end{equation}
where $\alpha$, $\nu$, k, E$_{g}$ and $\eta$ are the absorption coefficient, frequency, absorption constant, band gap energy and exponent of the equation, respectively \cite{Parhi_SSS_2008}. The value of the exponent $\eta$ depends on the type of the transition/band gap in the material and can have values of 1/2, 2, 3/2 and 3 for the direct allowed, indirect allowed, direct forbidden and indirect forbidden transitions, respectively \cite{Tauc_PSS_1966, Parhi_SSS_2008}. However, in the case of diffuse reflectance spectra, the $\alpha$ in Tauc plot can be replaced by the K-M function to estimate the band gap of the material \cite{Michalska2021}. As reported in \cite{HimanshuPRB21}, the LVNO samples has indirect band gap, therefore, the value of exponent, $\eta$ is 2, therefore, [F(R$_{\infty}$) $\times$ h$\nu$]$^{1/2}$ versus h$\nu$ is plotted for both the samples, shown in Fig.~\ref{Fig3}(c). This curve shows the linear and nonlinear portions which are the characteristics for allowed transitions. The linear portion of the curve which lies in the range approximately 3~eV to 4.5~eV characterizes the fundamental absorption of the materials and the non-linear portion of the curve correspond to a residual absorption, which involves defect states. The $x-$axis intercept along the optical absorption edge of the samples estimates the optical band gap, which is calculated as 3.2 and 2.7~eV for the 1000\degree C and 1250\degree C, respectively. Also, Sun {\it et al.} reported that monoclinic LaVO$_{4}$ phase possess indirect band gap of 3.5~eV, while tetragonal phase has direct band gap of 3~eV through the first-principle calculations \cite{SunJAP10}. Therefore, it can be clearly inferred that band gap decreases upon increasing the sintering temperature, as a result of increase in tetragonal phase in LVNO-1250 sample and further enhances the photoluminescence. This is also in agreement with the previously reported literature \cite{HimanshuPRB21, Michalska2021, Lopez_JSST_2012}. 

\begin{table}[h]
	\centering
	\caption{The peak fitting parameters for the observed vibrational frequencies ($\omega$$_{obs}$) of the LVNO-1000 and LVNO-1250 samples at room temperature in the wavenumber regions (50--1000 cm$^{-1}$).}
	\label{Raman_low}
	\begin{tabular}{p{2cm} p{3cm} p{3cm}}
		\hline	
		Sample & LVNO-1000 & LVNO-1250 \\
		\hline
		Peak & $\omega$$_{obs}$  & $\omega$$_{obs}$\\
		\hline
		\hline   
		P1  & A$_{g}$(63.8)  & A$_{g}$(64.2)   \\
		P2  & A$_{g}$(71.2)  & A$_{g}$(70.9)   \\
		P3  & B$_{1u}$(90.2) & B$_{1u}$(91.0)   \\
		P4  & B$_{g}$(103.5) & B$_{g}$(103.)    \\
		P5  & B$_{1g}$(116.7)& B$_{1g}$(115.9) \\
		P6  & B$_{g}$(127.8) & B$_{g}$(127.0)    \\
		P7  & E$_{g}$(138.4) & E$_{g}$(139.6)     \\
		P8  & A$_{2g}$(146.8) & A$_{2g}$(146.4)    \\
		P9  & B$_{g}$(159.8) & B$_{g}$(160.6)   \\
		P10 & A$_{g}$(173.4) & A$_{g}$(174.1)   \\
		P11 & A$_{g}$(189.8) & A$_{g}$(189.7)  \\
		P12 & B$_{g}$(209.8) & B$_{1g}$(211.9) \\ 
		P13 & B$_{g}$(219.9) &        \\
		P14 & A$_{g}$(232.8) &   \\
		P15 & B$_{2g}$(251.6) & B$_{2g}$(249.1)  \\
		P16 & B$_{g}$(285.6) &        \\
		P17 & A$_{1u}$(320.0) &         \\
		P18 & A$_{g}$(332.0) & A$_{1g}$(328.2)   \\
		P19 & E$_{g}$(347.7) & E$_{g}$(346.2)   \\
		P20 & A$_{g}$(373.6) & A$_{g}$(373.6)  \\
		P21 & B$_{g}$(396.9) & B$_{g}$(397.2)   \\
		P22 & A$_{g}$(423.4) &       \\
		P23 & B$_{g}$(439.9) & B$_{g}$(438.5)   \\
		P24  & 693.3  & A$_{g}$(710.3)    \\
		P25  & B$_{g}$(747.7)  & B$_{g}$(742.9)   \\
		P26  & A$_{g}$(770.0)  & A$_{g}$(769.9)   \\
		P27  & B$_{g}$(795.5)  & B$_{g}$(794.4) \\
		P28  & A$_{g}$(809.5)  & A$_{g}$(806.7)  \\
		P29  & A$_{g}$(821.1)  & A$_{g}$(820.0)  \\
		P30  & A$_{g}$(840.8)  & A$_{g}$(839.3)  \\
		P31  & B$_{g}$(858.8)  & B$_{g}$(858.8)   \\
		P32  & B$_{g}$(878.5)  & B$_{g}$(880.2)  \\
		\hline 
	\end{tabular}
\end{table}

Moreover, the Raman spectra recorded at a room temperature using a 532~nm excitation laser line for both the samples are presented in the Fig.~\ref{Fig4}(a-d). A Voigt profile function is employed to deconvolute and fit the observed Raman modes, as mentioned in the Table~\ref{Raman_low}. The intensity of observed Raman modes may fluctuate due to various factors, including the molecule's polarizability, the excitation wavelength of the laser source, and the concentration of the active groups \cite{Ashok_Jalcom_23}. The detailed description of the various types of Raman modes has been discussed in our previous works \cite{HimanshuPRB21, Ashok_Jalcom_23}. As per the group theory calculations reported, m-LaVO$_{4}$ consists of 72 vibrational modes (18B$_{u}$ + 18A$_{u}$ + 18A$_{g}$ + 18B$_{g}$), which includes 3 acoustic modes (A$_{u}$ + 2B$_{u}$), 33 infrared-active modes (16B$_{u}$ + 17A$_{u}$), and 36 Raman active modes (18A$_{g}$ + 18B$_{g}$) \cite{SunJAP10}. In this context, we employ Mulliken symbols, with notations A and B indicating that the vibrations are symmetric and anti-symmetric relative to the major axis of symmetry, respectively. The subscripts $g$ and $u$ denote that the vibrations are symmetric and anti-symmetric, respectively, concerning a center of symmetry. Meanwhile, the vibrational frequencies of tetragonal LaVO$_4$ (t-LaVO$_4$) consists of 36 vibrational modes and 33 zone-center optical phonon modes, with five $E_g$, four $E_u$, four $B_{1g}$, three $A_{2u}$, two $A_{1g}$, two $B_{2u}$, and one each of the $A_{1u}$, $A_{2g}$, $B_{1u}$, and $B_{2g}$ symmetry representations. Among these, the $A_{2u}$ and $E_u$ modes are infrared active, whereas the $A_{1g}$, $B_{1g}$, $B_{2g}$, and $E_g$ modes are Raman active. However, the remaining 
$A_{1u}$, $A_{2g}$, $B_{1u}$, and $B_{2u}$ modes are silent. The Raman, A$_{g}$, and B$_{g}$ modes can be experimentally identified by polarized Raman measurements. The detailed description of the modes can be referred to references in \cite{SunJAP10, Errandonea_JPCC_2016, Ashok_Jalcom_23}. 

The Raman modes in the lower wavenumber region for the LVNO-1000 and LVNO-1250 samples are presented as P0 to P23 and P24 to P32, respectively, in Fig.~\ref{Fig4}. Meanwhile, the corresponding details of Raman modes are described in Table~\ref{Raman_low}. In the LaV$_{0.5}$Nb$_{0.5}$O$_{4}$, the Raman peaks appear as a result of various vibrational bonds among La$^{3+}$, V$^{5+}$, Nb$^{5+}$ and O$^{2-}$. The Raman modes in the wavenumber region (600--800 cm$^{-1}$) are related to the stretching vibrations of O--(V/Nb)--O bonds and modes (300--450 cm$^{-1}$) correspond to the bending vibrations of O--V--O bonds and deformation/scissior modes of NbO$_{4}$$^{3-}$~\cite{Ashok_Jalcom_23}. The Raman modes in the low wavenumber region are associated with the translational lattice modes of La (due to its lower mass) and rotational modes of NbO$_{4}$$^{3-}$~\cite{SunJAP10}. Here, the P6 mode at 127.8 and 127.0 cm$^{-1}$ in LVNO-1000 and LVNO-1250 is linked with the translational motion of La atoms in the monoclinic phase. The P8 mode in the higher wavenumber region, located at 858.8 cm$^{-1}$ is the most intense mode, while the intensity of P6 mode enahnced with increase in temperature. 

\begin{figure*}
	\includegraphics[width=\linewidth]{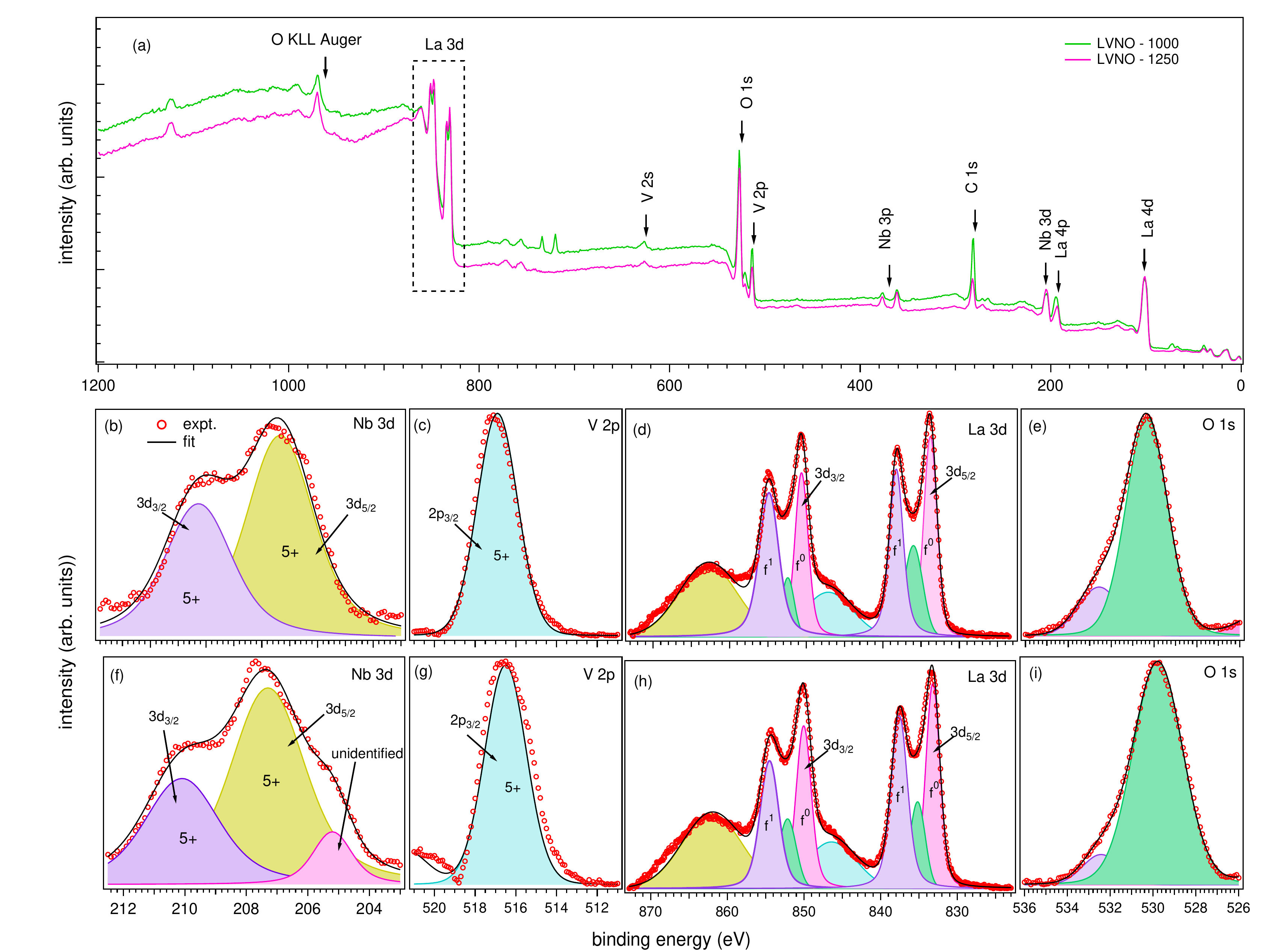} 
	\caption {(a) The room temperature XPS survey spectrum; the core level spectra of (b, f) Nb 3$d$; (c, g) V 2$p$; (d, h) La 3$d$ and (e, i) O 1$s$ elements of the LVNO sample sintered at 1000\degree C (b, c, d, e) and 1250\degree C (f, g, h, i), respectively.} 
	\label{Fig5}
\end{figure*}

Herein, the peak (P12) at 209.8 cm$^{-1}$ assigned to the A$_g$ mode corresponds to the torsional vibration of the VO$_4$ units, with the La atoms remaining nearly stationary. The peaks at 334 cm$^{-1}$ (P17) and 400 cm$^{-1}$ (P20) are attributed to A$_g$ modes associated with the bending vibrations of the O--V--O bonds. Furthermore, the peak observed at 839 cm$^{-1}$ (P30) corresponds to the A$_g$ mode, from the stretching vibrations of the O--V--O bonds. Furthermore, the presence of P8, P9, P10, P13, P14, P15, P17 and P18 modes confirms the existence of VO$_{4}$$^{3-}$ ion because none of these modes are visible in LaNbO$_{4}$ \cite{Errandonea_JPCC_2016}. In addition, there is an enhancement in the integrated intensity (P6) in the LVNO-1250, relative to LVNO-1000 [see Figs.~\ref{Fig4}(b, c)]. This behavior is related to the fact that the increase in the scheelite-tetragonal phase with increasing  sintering temperature. These monotonous effects on the
vibrational spectra are correlated to the substitution induced deformation of the VO$_{4}$$^{3-}$ tetrahedra and is clearly reflected in the FTIR spectra of the LVNO-1250 sample [see Fig.~\ref{Fig3}(a)]. For the LVNO-1000 sample, the effects correspond to the translational vibrations of La--O bonds and the rotational motions of NbO$_4^{3-}$ units. The peaks observed in the intermediate region (250--450~cm$^{-1}$) are associated with the O--V--O bending vibrations and the scissoring modes of NbO$_4^{3-}$. Meanwhile, in the high-wavenumber region (600--900~cm$^{-1}$), the Raman features arise from the stretching modes of Nb--O and V--O bonds. It can be clearly seen in Fig.~\ref{Fig4}(c) that the number of vibration bonds for LVNO-1250 sample decreases in comparison to the LVNO-1000 sample, which is in agreement with the lower coordination number (8) and the high symmetry ($D2d$) of the La$^{3+}$ ions in the tetragonal phase \cite{Jia_JPCB_2005}. In addition, Sun {\it et al.} also reported that Raman peaks for the tetragonal LaVO$_{4}$ are less as compared to the monoclinic LaVO$_{4}$ due to the high symmetry ($D2d$) of La$^{3+}$ ions, through the theoretical calculations \cite{SunJAP10}. 

\begin{figure*}[ht!]
	\centering
	\includegraphics[width=\textwidth]{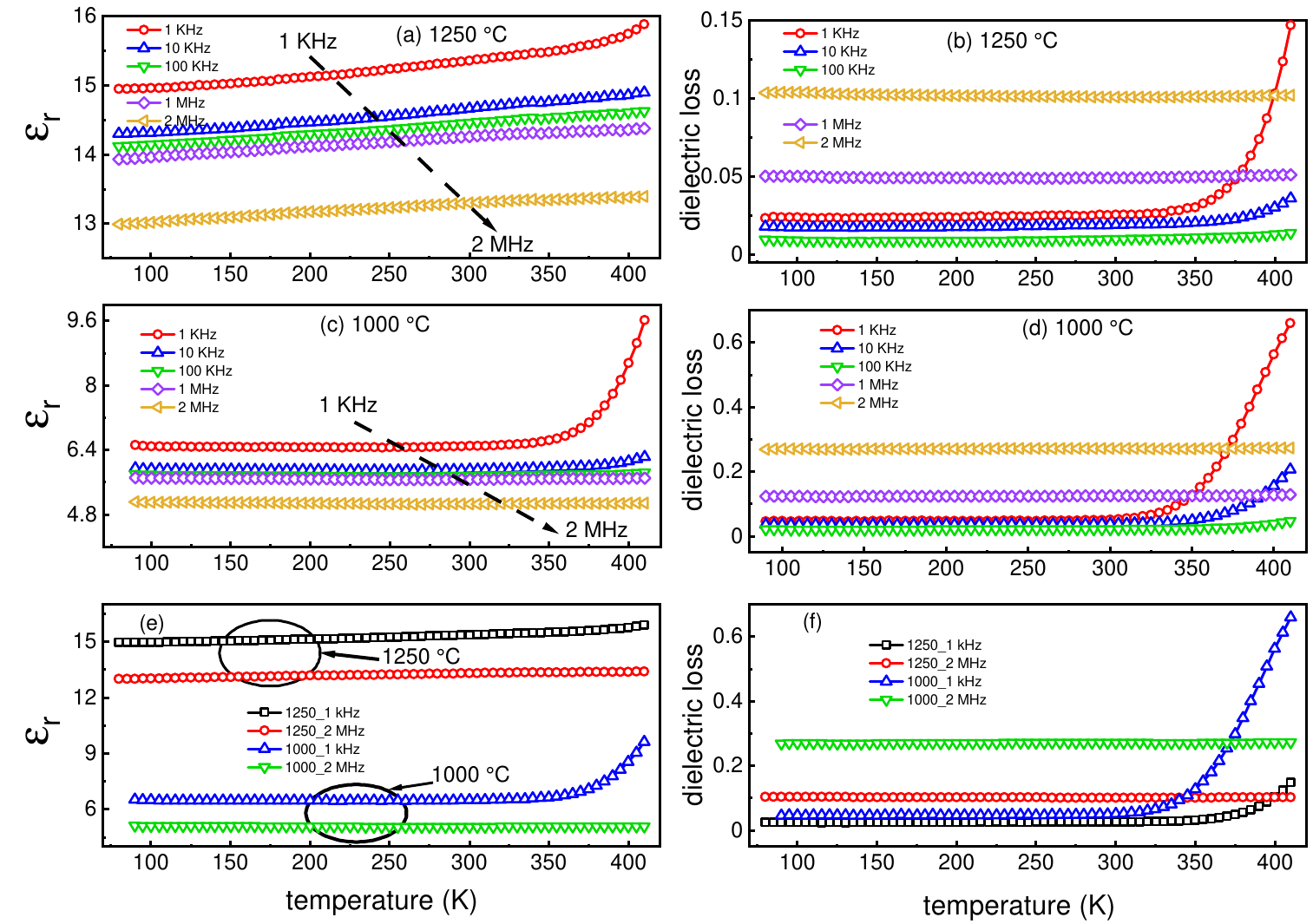}
	\caption{The variations of electric permittivity ($\epsilon$$_{r}$) and dielectric loss at various temperatures for selected frequencies for both the samples, sintered  1000\degree C and 1250\degree C temperatures; (a-d) the variations of electric permittivity and dielectric loss at various temperatures, (e-f) a comparison of electric permittivity and dielectric loss at two different frequencies of 1 kHz and 2 MHz, respectively. The arrow in this figure shows the direction of increasing frequency.}
	\label{ED}
\end{figure*}

Furthermore, the x-ray photoemission spectroscopy (XPS) is employed to probe the electronic structure by measuring the survey scan and specific elemental core-level spectra for both the samples. The detected peaks in the survey spectrum are assigned by their binding energies and exhibit strong correlation with the reported values \cite{Ashok_Jalcom_23}, as demonstrated in Fig.~\ref{Fig5}(a). The Nb~3$d$ core-level spectra for both the samples confirm the Nb$^{5+}$ oxidation state \cite{ShuklaJPCC19, ShuklaJPCC21}, as shown in Fig.~\ref{Fig5}(b, f). For  both the samples, the Nb~3$d_{5/2}$ and Nb~3$d_{3/2}$ peaks are observed at around 207.3~eV and 210.2~eV, respectively, with a spin--orbit splitting of around 2.9~eV~\cite{Onozato_JPCM_2016}. The O~1$s$ peak [see Fig.~\ref{Fig5}(e, i)] is observed at 529.8 eV (lattice oxygen) and the energy difference between Nb~3$d_{5/2}$ and O 1$s$ is 322.5 eV. The O~1$s$ peak, used as the binding-energy reference to compensate for charging effects, is centered at 530.2~eV, giving an energy difference of 322.9~eV between the Nb~3$d_{5/2}$ and O~1$s$ levels \cite{Atuchin_JESRP_2005}. There is an unidentified contribution at 205.2~eV, shown in pink color in Fig.~\ref{Fig5}(f). The energy difference of Nb~3$d_{5/2}$ and O 1$s$ is quite consistent, which reinforces Nb$^{+5}$ oxidation state in both the samples. 

Also, the V~2$p$ core-level spectra for both the samples in Figs.~\ref{Fig5}(c, g), the primary V~2$p_{3/2}$ peak occur at 516.9~eV, consistent with V$^{5+}$ oxidation \cite{Silversmit_JESRP_2004}. The La~3$d$ core-level spectra for both the samples exhibit the characteristic spin--orbit splitting (3$d_{5/2}$ and 3$d_{3/2}$) in Figs.~\ref{Fig5}(d, h). For the LVNO-1000 sample, the primary photoemission peaks are observed at 833.8~eV (3$d_{5/2}$) and 850.6~eV (3$d_{3/2}$), corresponding to a spin--orbit separation of 16.8~eV \cite{Mullica}, confirming the La$^{3+}$ oxidation state \cite{Guigoz_IJHE_2020}. However, a double peak structure is observed in the spin-orbit component, denoted by f$^{0}$ and f$^{1}$, corresponding to the 3$d^{9}$4$f^{0}$ and 3$d^{9}$$f^{1}$L configurations, shown in Figs.~\ref{Fig5}(d, h). Here, the L signifies the hole in the O 2$p$ valence orbital and arises from the electron transfer from the ligand valence band to empty 4$f$ orbitals \cite{ShuklaJPCC19, 39, Ravi_JALCOM_2018}. In addition to the La 3$d$ main peaks, broad satellite features at 847.9 and 862~eV are observed due to the plasmons \cite{Ashok_Jalcom_23}. For the LVNO-1250 sample, a small shift of 0.6 eV in the main La peaks toward lower binding energy is observed. The primary peaks appear at 833.2~eV (3$d_{5/2}$) and 850.0~eV (3$d_{3/2}$), with a spin--orbit splitting of 16.8~eV, while the corresponding satellite peaks occur at 837.6~eV and 854.5~eV. The energy separation between the primary and satellite peaks, which reflects the degree of metal--ligand orbital overlap \cite{Guigoz_IJHE_2020}, is calculated to be around 4.45~eV for both the samples. 

\begin{figure*}
	\centering
	\includegraphics[width=\textwidth]{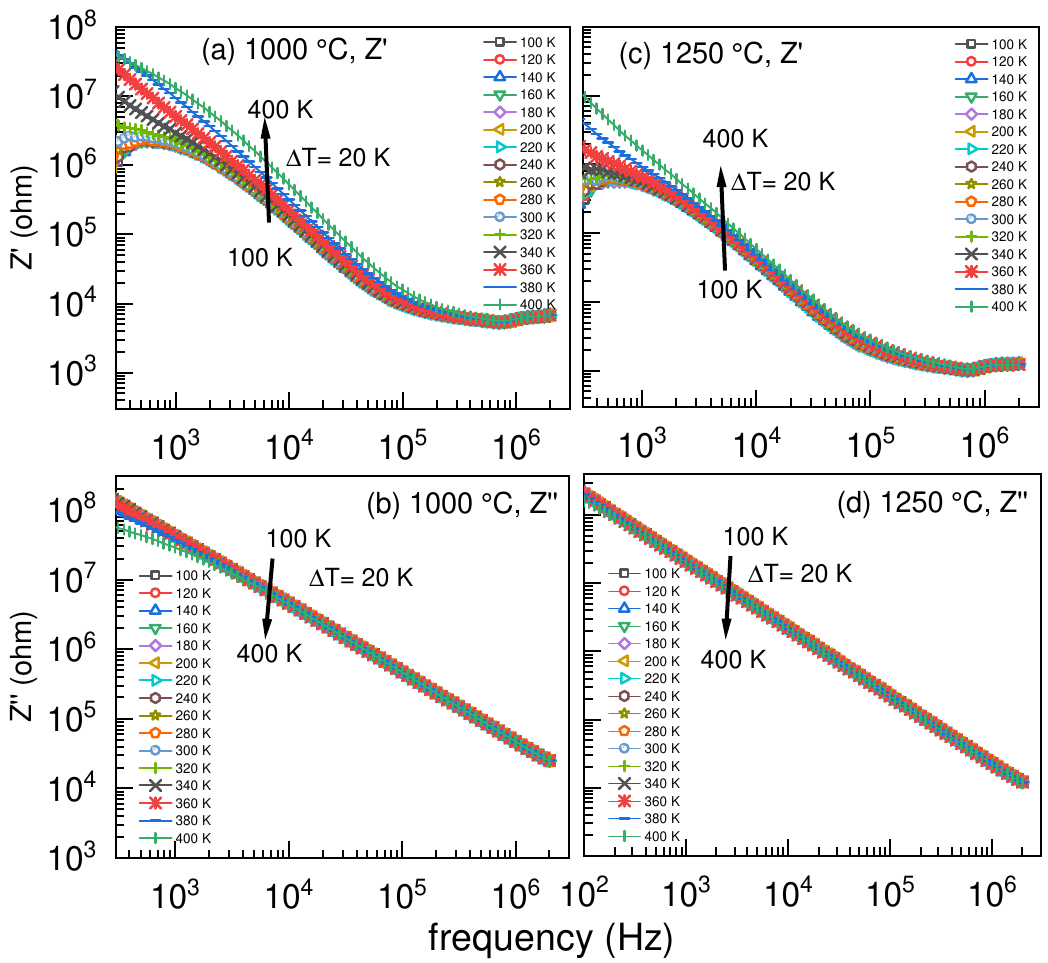}
	\caption{The real (Z$^{'}$) and imaginary (Z$^{''}$) parts of total impedance (Z), (a, b) for the LVNO-1000 and (c, d) for the LVNO-1250 samples. The arrow shows the direction of increasing temperatures.}
	\label{Z'Z''}
\end{figure*}

Finally, the temperature dependence of dielectric permittivity and dielectric loss of the LVNO samples are shown in Fig.~\ref{ED}. The results show a decrease in dielectric permittivity ($\epsilon$$_{r}$) with increasing frequency and an increase in permittivity with increasing temperature for both the samples, as shown in Figs.~\ref{ED}(a, c). The frequency-dependent behavior of the LVNO samples at selected temperatures can be explained using the space charge polarization mechanism \cite{RM_CI_22, Balaraman_JPD_22, Kesntini_APA_20}. According to this model, at lower frequencies, the charge carriers follow the applied field, which produces a higher amount of polarization, leading to a higher permittivity. The sample sintered at 1250\degree C exhibit higher permittivity compared to the one sintered at 1000\degree C, as a higher sintering temperature leads to greater densification and larger grain size of the ceramics, which improves the accumulation of charge carriers at the interface. The higher sintering temperature, accompanied by an increase in grain size, results a decrease in insulating grain boundaries, leading to an increase in polarization and, consequently, dielectric permittivity \cite{Sun_CI_15, Chandrakala_CI_16}. The higher amount of polarization gives rise to a large permittivity, as observed for the sample  sintered at 1250$^{\circ}$C \cite{Kumar_JMS_21, Alkathy_CI_16}. The temperature-dependent variations in permittivity are explained in terms of thermal activation of charge carriers. As the temperature is increased at a fixed frequency, the thermal energy associated with the dipoles helps them to orient along the direction of the applied field; hence, the dielectric permittivity is increased with an increase in temperature.  The total dielectric permittivity is the combined effect of the applied frequency and measured temperature. A similar type of temperature-dependent behavior is also observed for the dielectric loss data, as shown in Figs.~\ref{ED}(b, d). The lower dielectric loss at higher sintering temperature is attributed to the greater densification (as porosity leads to dissipating the energy in dielectric materials) of the ceramics and a decrease in oxygen vacancies. An increase in dielectric loss with temperature is due to an increase in space charge polarization, which leads to an increase in conductivity of the ceramic samples, hence an enhancement in the dielectric loss. The dielectric loss at high temperature is mainly attributed to the conduction loss. The frequency-dependent dielectric loss data show an increase in dielectric loss above 100 kHz, indicating that the resonance frequency for the LVNO molecule is above 100 kHz. The comparison of dielectric permittivity Fig.~\ref{ED}(e) and dielectric loss Fig.~\ref{ED}(f) shows a higher permittivity and lower dielectric loss for sample sintered at 1250\degree C.

Moreover, the variation of real (Z$^{'}$) and imaginary (Z$^{''}$) parts of total impedance at selected temperatures is shown in Fig.~\ref{Z'Z''}. The impedance data are normalized by the geometrical correction factor $\frac{A}{d}$, where A is the total electrode area and $d$ is the thickness of the sample. The real part of impedance (Z$^{'}$) is categorized into two: (a) frequency-dependent data at lower frequencies related to the $a.c.$ conduction in the material gives a rapid decrease in the impedance due to the interfacial polarization, and (b) merging of all curves with each other indicates the possible release of space charge due to reduction in potential barriers. The rapid decreasing behavior of the real impedance with frequency is due to an enhancement in the mobility of charge carriers  \cite{Javed_MRB_23}. The analysis shows that the real part of the total impedance (Z$^{'}$) increases with temperature, indicating the enhancement in the resistive nature of the LVNO samples, see Figs.~\ref{Z'Z''}(a, c). The imaginary part (Z$^{''}$) decreases with temperature or nearly remains the same with temperature, indicating no variations in the capacitive nature of the samples, see Figs.~\ref{Z'Z''}(b, d). The higher values of impedance at lower frequencies are due to interfacial polarization, where the capacitive nature of the samples dominates according to the time constant and temperature of the samples. The sample sintered at 1250\degree C show the lower values of real impedance (Z$^{'}$), indicating the better crystalline nature of the sample having the lower porosity with larger grain size \cite{Zhang_MRB_22}, as supported by the dielectric permittivity and loss data shown in Fig.~\ref{ED}.

\begin{figure}
    \centering
    \includegraphics[width=0.48\textwidth]{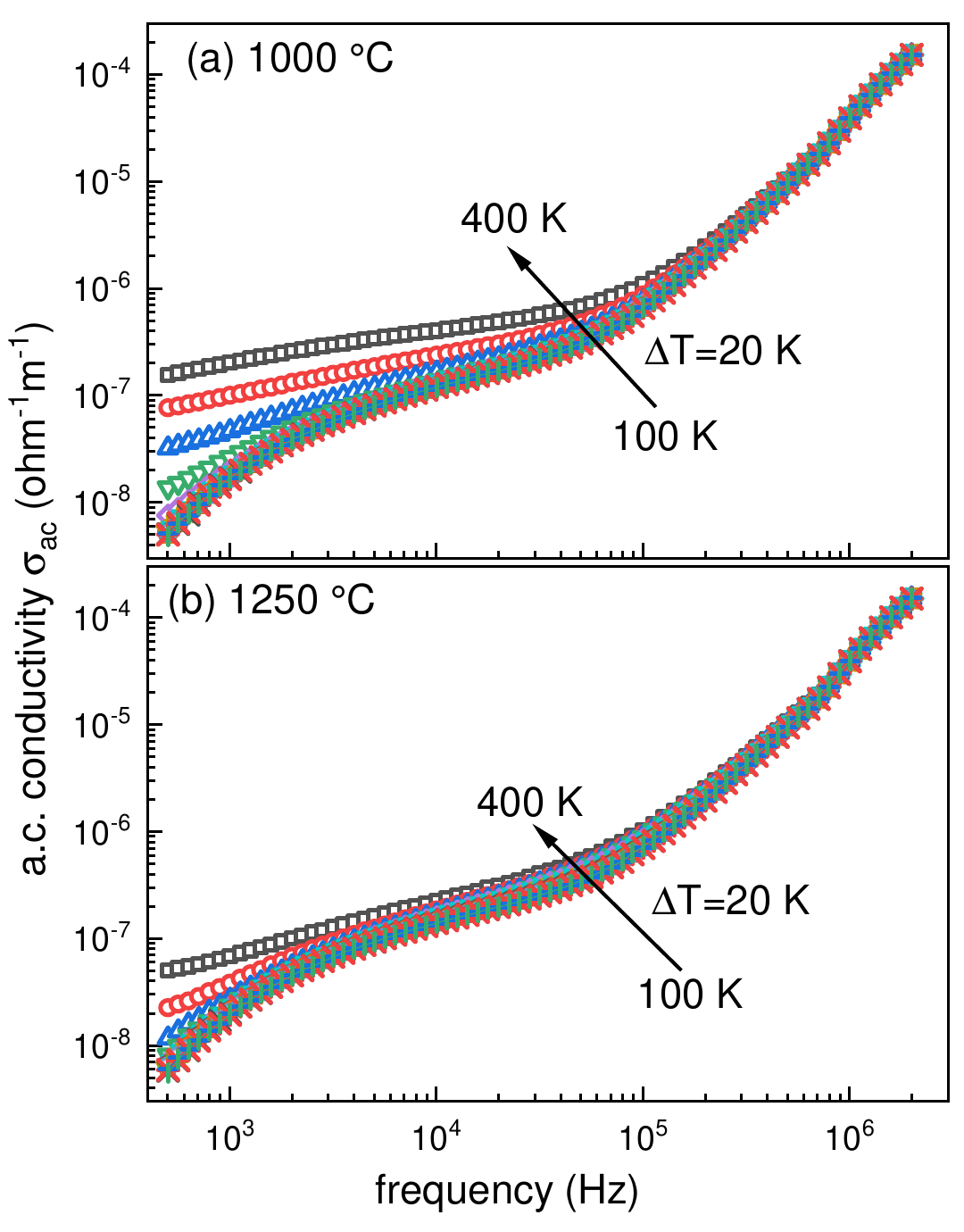}
    \caption{The variation of ac conductivity with frequency for (a) 1000$^{\circ}$C and (b) 1250$^{\circ}$C sintered samples. The arrow shows the direction of increasing temperature.}
\label{ACS}
 \end{figure}

Note that the {\it a.c.} conductivity is a response of the carriers in phase with the applied electric field, which depends on the frequency of the applied field and temperature. It is a combined response of relaxation, localized hopping and diffusion of charge carriers. The variation in conductivity with frequency is a crucial tool for determining the type of carriers that dominate in the total conductivity. If the conductivity increases (decreases) with an increase in frequency, the conductivity is dominated by the contributions from the bound (free) charge carriers \cite{RM_CI_22, Sumi_JAP_10, Meena_PB_24, Meena_Small_25}. In an alternating electric field, the charge carriers move from one state to another by a hopping conduction. At lower frequencies, carriers undergo long-range hopping due to the considerable time period available to them; in contrast, a shorter time period at higher frequencies results in short-range hopping. The hopping probability increases with temperature due to an increase in the activation energies of carriers as well as a reduction in barrier height, resulting in an increase in conductivity. The $a.c.$ conductivity, depends also on the sample dimensions, can be calculated using the following relation \cite{Meena_PB_24, Raut_JAP_11, Meena_Small_25}:  
\begin{equation}
\sigma_{ac}(\omega)= \frac{t}{A} \left[ \frac{Z'}{Z'^2+Z''^2} \right]
\label{ACS1}
\end{equation}
Here, $t$ is the thickness of the dielectric material, $A$ is the total area of the metal electrodes, $Z'$ and $Z''$ are the real and imaginary impedance of the total impedance $Z$.

Further, the $a.c.$ conductivity calculated using equation \ref{ACS1} is shown in Fig.~\ref{ACS} at selected temperatures. The conductivity is found to increase with an increase in temperature due to the thermal activation of charge carriers; however, the conductivity remains in a similar order for samples sintered at 1000\degree C and 1250\degree C. It is found that conductivity increases with frequency for both the samples, confirming the contributions from the bound charge carriers. The conductivity data show two different types of regions: (a) a plateau-like region at lower frequencies and (b) dispersive type behavior at higher frequencies. This type of frequency-dependent variation in conductivity is explained on the basis of the Funke jump relaxation model \cite{Funke_PSSS_93}. According to this model, charge carriers follow long-range motion at lower frequencies, resulting in a frequency-independent contribution. On the other hand, at higher frequencies, the dispersive nature is due to the competition between successful and unsuccessful hopping due to the shorter time period available with carriers. The analysis shows that the discrete nature of conductivity at lower frequencies is due to the accumulation of space charge over the potential barriers, while the merging of the conductivity data at higher frequencies for both the samples confirm the reduction of potential barriers at higher frequencies due to the possible release of space charge. The sample sintered at 1000\degree C exhibits a large dispersive nature, as small grain size and lower densification result in a large potential barrier, thereby increasing the ratio of unsuccessful hopping to successful hopping. On the other hand, sample sintered at 1250\degree C exhibit a larger grain size and higher densification, resulting in a higher amount of successful hopping over potential barriers, thus a decrease in dispersive behavior is observed. The analysis also shows that the dispersive nature shifts towards the higher frequency side with an increase in temperature for both the samples, which can be explained by the enhancement in thermal activation energy available with the carriers. At higher frequencies, lesser time scale leads to the unsuccessful hopping of carriers towards neighboring sites. As the temperature increases, the average thermal energy available to carriers is enhanced, allowing long-range hopping even for the shorter periods available at higher frequencies. The long-range hopping extends the dispersive nature towards the higher frequency side with an increase in temperature for both the samples \cite{Sumi_JAP_10, Rao_JSNM_20, Moualhi_JALCOM_22}.

\section{\noindent ~Summary and Conclusions}

The LaV$_{0.5}$Nb$_{0.5}$O$_4$ samples were successfully synthesized via the conventional solid-state reaction route. A pronounced dependence of the physical and dielectric properties on the sintering temperatures 1000\degree C and 1250\degree C has been observed. The room-temperature Rietveld refinement of the powder X-ray diffraction patterens reveal an increase in the tetragonal phase fraction to 96\%, accompanied by a minor monoclinic phase fraction of 4\%, upon sintering at 1250\degree C. The FE-SEM analysis illustrates the surface morphology, which indicates a clear increase in particle size with increasing sintering temperature. The high-resolution transmission electron microscopy further confirms the crystallinity and coexistence of mixed phases in both the samples. The Raman and Fourier transform infrared spectroscopy measurements validate the presence of the characteristic vibrational modes. The UV–vis diffuse reflectance spectroscopy indicates an optical band gap of 3.2~eV for the sample sintered at 1000\degree C; however, as the tetragonal phase becomes more prominent at 1250\degree C, the band gap narrows to 2.7~eV, which is advantageous for luminescent applications. The x-ray photoemission spectroscopy reveals the electronic structure and confirms the valence state of the constituent elements. Lastly, the dielectric studies reveal that the sample annealed at 1250\degree C exhibits higher dielectric permittivity ($\epsilon_r$) and lower dielectric loss. These results clearly demonstrate that the sintering temperature plays a crucial role in tailoring the physical, electronic, and dielectric properties of LaV$_{0.5}$Nb$_{0.5}$O$_4$ orthovanadate. \\

%\section*{\noindent ~Author contributions}

%All authors have contributed to this work.

%\section*{\noindent ~Declaration of competing interest}

%The authors declare that they have no known competing financial interests that could have appeared to influence the work reported in this paper. 

%\section*{\noindent~Data Availability}

%The data that support the findings of this study are available upon reasonable request. 

\section*{\noindent ~Acknowledgments}
We acknowledge Physics Department at IIT Delhi for the XRD and Raman Spectroscopy measurements. The authors thank Gaurav Gupta, Vikas Joshi and Jiten Dhaka for their help during synthesis and analysis of the samples. We thank Central Research Facility (CRF) for the FESEM, EDX, HR-TEM, UV-vis DRS, FTIR and XPS measurements. The samples are synthesised in a high-temperature furnace (Nabertherm GmbH, Germany), funded by BRNS through the DAE Young Scientist Research Award (Project Sanction No. 34/20/12/2015/BRNS). RSD acknowledges SERB--DST for the financial support through a core research grant (project reference no. CRG/2020/003436).  \\

\end{document}